\documentclass[manuscript=article]{achemso}

\usepackage[T1]{fontenc} 
\mciteErrorOnUnknownfalse

\author{Subhadip Ghosh}
\author{Arun Roy}
\email{aroy@rri.res.in}
\affiliation[A shared footnote]
{Raman Research Institute, C.V. Raman Avenue, Sadashivanagar, Bengaluru 560080, India}

\title{Optical and structural properties of starch films formed 
on drying droplets of gelatinized starch solutions}

\abbreviations{}

\begin{document}


\begin{abstract}

We report the optical and structural properties of 
starch films formed on drying droplets of starch 
solutions on a flat substrate. 
The starch films formed after drying 
are circular in shape and have an  
azimuthally symmetric "M-shaped" height
profile along their diameter.
These films are found to be semi-crystalline 
in nature and transparent in visible light.  
Our experimental results show
that these films are optically 
uniaxial at their center but biaxial 
away from the center. The variation of 
birefringence along the radial direction 
is studied for the films formed after drying 
droplets with different volumes and starch concentrations.
The cryo-SEM studies on the 
drying droplets and the films reveal 
the micro-structures and  
the possible origin 
of birefringence of these films.

\end{abstract}

\section{Introduction}
The studies on various properties of solute films formed on drying of
solution droplets casted on a flat substrate is an 
active field of research during last two decades.
The solution droplets of various solutes
such as polymers 
\cite{doi2009, doi2008, doi2006}, 
DNA \cite{smalyukh2006, maheshwari2008}, 
nanotubes \cite{duggal2006, zhang2010, li2006, zhao2015}, 
nanorods \cite{querner2008, martin2014}, 
liquid crystals \cite{yodh2017, chu2018}, 
colloids \cite{deegan1997, deegan2000, parisse1997, haw2002}, 
surfactants \cite{jing2021}, 
salts \cite{basu2020, choudhury2013} 
and biological materials 
\cite{brutin2011, gorr2013, pal2020} 
have been used for this purpose. The most
spectacular phenomenon found after drying of such a 
dilute solution droplet is the coffee ring effect \cite{deegan1997, deegan2000}. 
In this phenomenon a circular stain is found to form
due to the capillary flow of 
solute particles towards the pinned 
edge of the droplet. Interestingly, 
self-assembled structures of solute 
ingredients have also been found in some cases which makes 
the deposited thin films birefringent 
\cite{smalyukh2006,li2006,zhao2015,yodh2017,beyer2012}. 
It is found that the droplets
with high concentration of polymer and high contact angle with
the substrate produce films with distorted surface profiles after
drying \cite{doi2006, pauchard2003, pauchard20032nd, pauchard20033rd}. 
In these cases, an elastic or 
glassy but solvent permeable skin forms on top of the droplets 
during drying. 
Further drying decreases the enclosed droplet 
volume which buckles the skin and produces varieties of surface 
distortion in the deposited films.    

Starch, the most abundant and consumable 
carbohydrates, is the source of 
large polysaccharides found in nature. 
Starch naturally occurs in the form of grains and
is made of amylose and amylopectin which are 
generally linear and branched polymers respectively.
The starch grains on heating in excess water 
swell and rapture irreversibly by imbibing 
water above a certain temperature. This process allows 
the starch polysaccharides to disperse in water and is known as
gelatinization of starch \cite{wang2013}. The gelatinized 
starch solutions casted on flat substrate produce transparent 
biodegradable starch films after 
drying \cite{kim2020, navid2021, dawei2019}.
In recent years, 
starch based bio-plastics have attracted much attention as a 
replacement of synthetic polymer films in 
plastic industries \cite{vieira2011}. 

In this report, we have studied the 
properties of the biodegradable potato 
starch films formed after drying droplets of  
gelatinized starch solutions casted on a plastic 
substrate. We have investigated the drying dynamics 
of these droplets and find that the circular films produced after drying
are linearly birefringent. The films are 
optically uniaxial at the center with an optic axis normal 
to the film whereas they are biaxial away 
from the center. The polarised optical 
microscopy studies carried out on these films
establish that the three principal axes of the 
indicatrix in the biaxial region of the circular films are 
along  the radial, azimuthal and perpendicular 
directions respectively. We measure the variation of 
linear birefringence along the diameter of 
these films which provides valuable 
insight on the orientational order of starch bio-polymers 
developed on drying. The SEM studies on drying droplets are
performed to investigate the micro-structures 
inside the droplets and the films. Based on these 
experimental results, a possible mechanism of the acquisition of
optical anisotropy observed in these starch films is proposed.

\section{Experimental}

\textbf{Extraction of potato starch:}
The starch grains were extracted from the potatoes 
procured locally from the market.
The washed potatoes were peeled and then
grated gently followed by mixing with 
excess water at room temperature. This 
mixture was stirred properly and then the dispersed starch grains 
in water were separated from the remnants of the grated potatoes by using a sieve. 
The dispersed starch grains being of slightly higher density 
than water precipitated after few minutes.
After complete precipitation, 
the excess water on top of the sediment was 
decanted and the sedimented starch content was dried in air 
at room temperature for 3--4 days.   
The potato starch sample after 
complete drying appears as white powder.

\paragraph{\textbf{Preparation of starch solutions and films:}}
A beaker containing starch grains dispersed 
in water was closed tightly by using 
aluminum foil and kept at 90$^\circ$C for 1.5~hours with continuous staring 
of the sample using magnetic starrier. At this temperature, the starch 
grains become completely gelatinized. The cooled sample was then 
homogenized using a ultrasonic probe sonicator (QSONICA Q700). 
The 17 ml of the starch solution was taken in a 20 ml glass vowel and 
sonicated in pulsed mode (2~sec./4~sec. on/off) for 10 minutes with 40\% 
amplitude of vibration using a 6 mm diameter tip. The sonicated starch 
solution was then cooled to room temperature by 
keeping it in a water bath for 5 minutes.
In this way, five different gelatinized starch solutions were
prepared with concentrations 10.8 wt\%, 
9.5 wt\%, 8.0 wt\%, 6.5 wt\% and 5.0 wt\% respectively. 
Different volumes of these solutions 
were dropcasted on flat plastic substrates by using a micro-pipette. 
The droplets on complete drying at room 
temperature and ambient humidity
produce the starch films on the substrate.
The flat plastic substrates were used as the films produced can be easily
detached from the substrate for further studies. 

\paragraph{\textbf{Weight measurement during drying:}}
The weight loss during drying of a starch solution 
droplet was measured using a quartz crystal 
micro-balance with precision of $10^{-4}$~g.
The 150 $\mu$l of starch solution was dropcasted on the flat surface 
of a plastic petridish using a micro-pipette.
This droplet was then allowed to dry in air
at room temperature (about 25$^\circ$C) and humidity in the
range 53\%--55\%. The petridish was kept on the micro-balance and 
the weight readings were taken in intervals of
5 minutes till complete drying of the droplet. The weight of the
droplet was calculated by subtracting the weight of the empty petridish
taken prior to the dropcasting.
The measurements were performed for 150 $\mu$l droplets 
with different initial starch concentrations.

\paragraph{\textbf{Polarising optical microscopy 
(POM) and UV-Visible spectroscopy:}} 
Both the conoscopic and orthoscopic POM studies of starch films were
performed by using Olympus BX50 polarising optical microscope. 
The UV-Visible transmission spectroscopy 
studies of starch films were performed 
by using Lamda 35 Perkin Elmer spectrophotometer.  

\paragraph{\textbf{Measurement of linear birefringence:}}
A He-Ne laser beam (5~mW) was passed through
a polariser with its pass axis at 0$^\circ$, a photo elastic 
modulator (PEM-100, Hinds Instruments)
with its axis set at an angle of 45$^\circ$ to the polariser, the sample
starch film and the analyser oriented at 90$^\circ$ to the polariser
(see fig.~S1 in supporting information). 
Then the
intensity of the laser beam was monitored by using a photodiode. 
The radial direction of the circular starch film was set 
parallel to the the axis of the PEM during the measurements.
The voltage signal generated by the photo diode is
\begin{equation}
\label{eqn:voltage}
V=\frac{V_0}{2}[ 1-cos(\delta _R -\delta)]
\end{equation}
where $V_0$ is the voltage signal corresponding to the light
intensity after polariser and $\delta$, $\delta _R$ are 
the linear optical phase retardations introduced by the starch 
film and PEM respectively. The phase retardation
$\delta _R = A_0cos(\omega t)$, where $A_0$ and $\omega$ are the amplitude
and frequency of sinusoidal variation of retardation in PEM. The values
of these parameters were set as 2.405 radian and 50 KHz respectively.

The eqn.~\ref{eqn:voltage} can be expanded in Fourier series as,
\begin{equation}
\label{eqn:fourier}
V=V_{DC}+V_0[-J_1(A_0)\sin\delta\cos(\omega t)
+J_2(A_0)\cos\delta\cos(2\omega t)+ \textrm{higher~harmonics}]
\end{equation} 
where $V_{DC}$ is the DC part of the signal 
and $J_1(A_0)$, $J_2(A_0)$ are Bessel 
functions of first kind of order 1 and 2 respectively. 
The AC part of the signal was separated by
a signal conditioning unit (SCU-100, Hinds Instruments) and 
the rms amplitudes of first ($V_{1f}$) and second 
($V_{2f}$) harmonics of it were measured by a lockin amplifier (SR 830). 
Then from eqn.~\ref{eqn:fourier}, the retardation of the film  can be obtained as,
\begin{equation}
tan\delta = \frac{J_2(A_0)}{J_1(A_0)}\times \frac{V_{1f}}{V_{2f}}
\end{equation}
where $J_1(A_0)=J_1(2.405)=0.5191$ and $J_2(A_0)=J_2(2.405)=0.4317$.

The thickness profiles of the circular starch films were measured by
cutting them along a diameter. The images of the cross-section
of the films were taken
by using a microscope (Olympus BX50) attached with a digital camera 
(Canon EOS 80D). These images were then analyzed by using a digital image
processing software (imageJ, NIH) to obtain the thickness 
profiles of the films along the diameter.
The linear birefringence
along the diameter of a film was calculated 
from the ratio of the measured optical path difference and 
the corresponding thickness.

\paragraph{\textbf{Cryo-SEM studies:}}
The droplets of volume 300 $\mu$l with 9.5 wt\% starch concentration were 
dropcasted on a plastic petridish and left to dry in air. 
These droplets generally take about 18 to 20 hours 
to dry completely at about 25$^\circ$C and 45\% - 50\% humidity.
The liquid nitrogen was poured in the plastic 
petridish to freeze the droplets at different times after
dropcasting to study the micro-structures. 
A frozen droplet was then removed from the plastic 
petridish and immediately fixed vertically on an aluminium sample holder by 
using colloidal graphene conducting paste such that 
half of the droplet protrudes out of the holder (see fig.~S2 in 
supporting information).
The sample with the holder was again
dipped in liquid nitrogen and then transferred to the sample stage
of the cryo-preparation chamber 
(Quorum Technologies PP3000T) kept at -170$^\circ$C and 
the pressure maintained in this chamber was 10$^{-4}$ to 10$^{-5}$ mbar.
The sample was then heated to -90$^\circ$C and kept at this temperature
for 5 minutes to make it soft for cutting through its diameter.
After cutting, the newly opened cross section of the sample was
left for 15 minutes at this temperature 
and pressure for ice sublimation followed by 
sputter coating it with platinum for 60 seconds. 
Finally, the sample was transferred to the microscope 
stage of the field emission scanning electron microscope 
(CARL ZEISS system, ULTRA PLUS model) 
which was kept at -190$^\circ$C and the 
pressure maintained in the sample chamber was about
10$^{-6}$ mbar.

\paragraph{\textbf{X-ray diffraction (XRD) studies:}}
A DY 1042-Empyrean (PANalytical) x-ray diffractometer 
with PIXcel 3D detector was used for acquiring
the XRD profiles of the samples using 
CuK$_\alpha$ radiation of wavelength 1.54 \AA{}.
The samples were kept 
on a flat silicon stage of the diffractometer 
to acquire their XRD profiles and 
the measurements were performed 
in the grazing angle of incidence of the x-ray
beam. The silicon stage gives a flat XRD profile which does 
not interfere with sample profiles.
 
The XRD profiles of solution 
droplets during their drying were taken after dropcasting a 
150 $\mu$l droplet with 9.5 wt\% of initial starch 
concentration on a glass plate. The XRD measurements were performed 
in the grazing angle of incidence of the x-ray
beam. The baseline corrections  
were made by subtracting the profile 
of the empty glass plate from the sample profiles. The time interval
between two successive measurements was 30 minutes.

\section{Results and discussion}
The native potato starch grains are spherical or elliptical in shape
with size from about 10~$\mu$m to 60~$\mu$m. The native starch grains 
are semi-crystalline in nature and possess birefringent 
property.
The potato starch grains on heating in excess water 
swell and ultimately disrupt on absorbing water in the 
temperature range from 60$^\circ$C to 75$^\circ$C.
This irreversible transformation phenomenon is known as starch 
gelatinization \cite{wang2013}. In this process, starch 
grains loose its crystalline and 
birefringent property \cite{jenkins1998,liu1991}.
The resulting gelatinized starch solution is 
inhomogeneous due to the presence of so 
called "ghost" structures of grains and the entangle 
starch bio-polymers \cite{zhang2014,debet2007}. 
A homogeneous starch solution can be prepared by 
sonicating the gelatinized sample.
Probe sonication of this sample breaks the 
starch bio-polymers and gives rise to a homogeneous,
transparent and less viscus starch solution. These starch solutions with
various initial starch concentration were dropcasted on plastic substrate
and the circular transparent potato starch films were formed on drying 
in ambient condition. 

\begin{figure}[H]    
    \includegraphics[width=10cm]{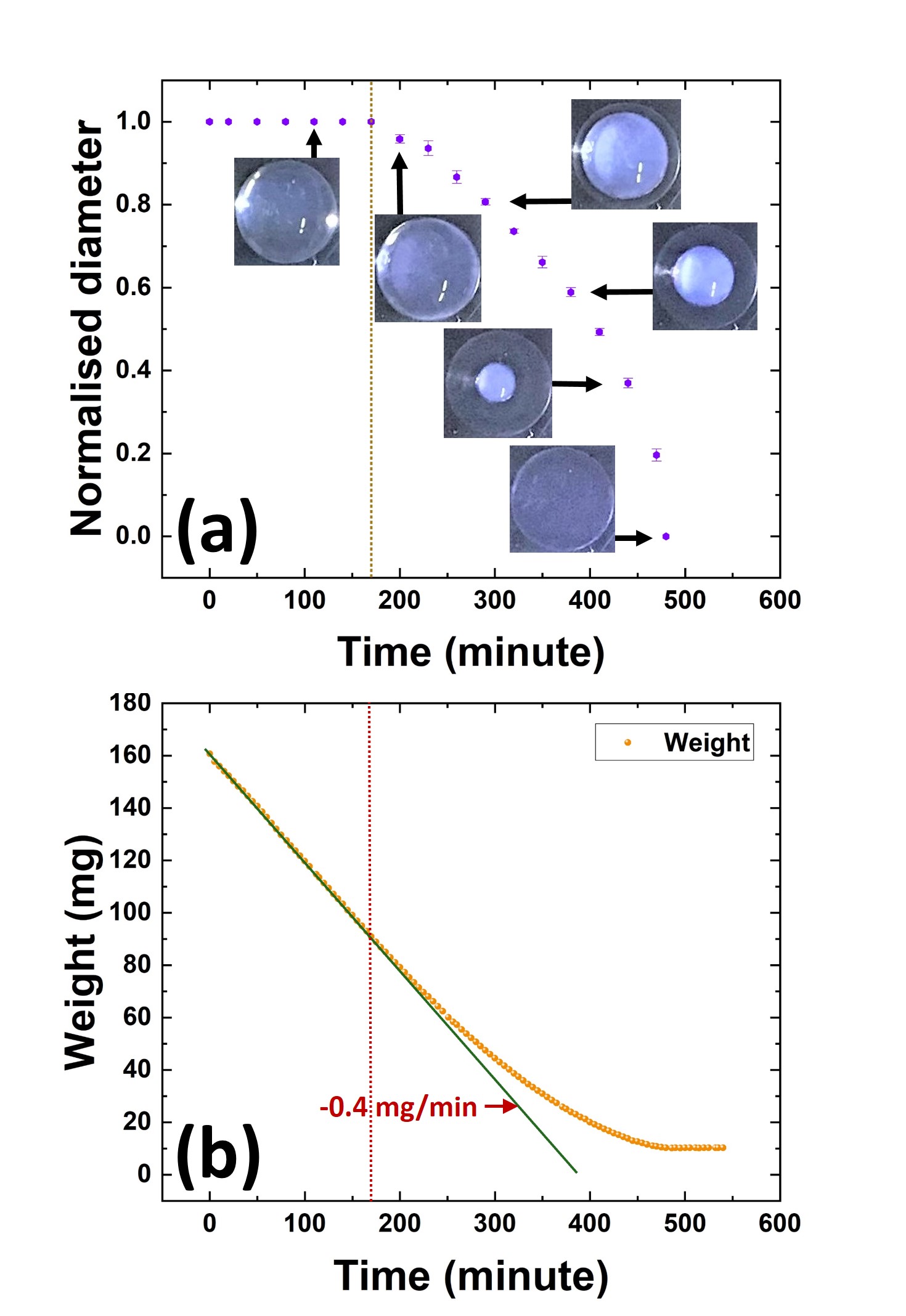}
    \centering
    \caption{(a) The variation of normalized diameter of a 150~$\mu$l 
    starch solution droplet with
    5.0~wt\% starch concentration during its drying. The images show 
    the sample corresponding to some of the data points as indicated. 
    (b) The variation of weight during its drying. The straight line
    (green) shows the initial linear variation of the weight. 
    }
    \label{fig1}
\end{figure}

The drying dynamics of a starch droplet on a plastic substrate
was studied as a function of time after its dropcasting.
Fig.~\ref{fig1}a shows the images of a 150 $\mu$l droplet with 
5 wt\% starch concentration at different time intervals of its drying.
Initially, the droplet dries with its edge pinned to the substrate.
After about 170 minutes, the droplet edge starts to recede towards
the center with the formation of a peripheral film from the initial
pinned edge of the droplet.  Further drying reduces the effective 
diameter of the central spherical cap of the sample.
The time variation of the diameter of the circular 
edge of the droplet cap normalized by the initial diameter of the 
droplet is shown in
fig.~\ref{fig1}a. Initially, this normalized diameter remains 
constant and starts to decrease after 
about 170 minutes. The complete
drying of this droplet gives rise to a circular starch film
with diameter same as that of the initial droplet.

The variation of the weight of this 
drying droplet 
with time is shown in 
fig.~\ref{fig1}b.
Initially, the weight of the droplet decreases 
linearly due to constant rate of evaporation of water 
which was about 0.40 mg/min. 
The rate of weight loss starts to
decrease gradually after about 170 minutes till 
the complete drying of the droplet. Therefore, we find that the 
decrease in the rate of weight loss is correlated with the receding 
of the droplet edge.
The sample has negligible weight loss after its complete drying.

The similar dynamics was also found for other
starch solution droplets with same initial volume 
but different starch concentrations. The time variation of 
weight for a 150 $\mu$l starch solution droplet 
with 9.5 wt\% of initial starch concentration is shown in
fig.~S3 of the supporting information. Initially the weight loss 
rate was constant at about 0.36 mg/min and the rate started to decrease subsequently with
the receding of the droplet 
edge which occurred above 100 minutes from its dropcasting.     
But our SEM studies reveal that 
a gel network forms in the droplet of this concentration within 
about 10 minutes of its dropcasting. 
Therefore, the decrease in the
weight loss rate is independent of sol-gel transition but
depends on the drying dynamics of the droplets.

\begin{figure}[t!]    
    \includegraphics[width=14cm]{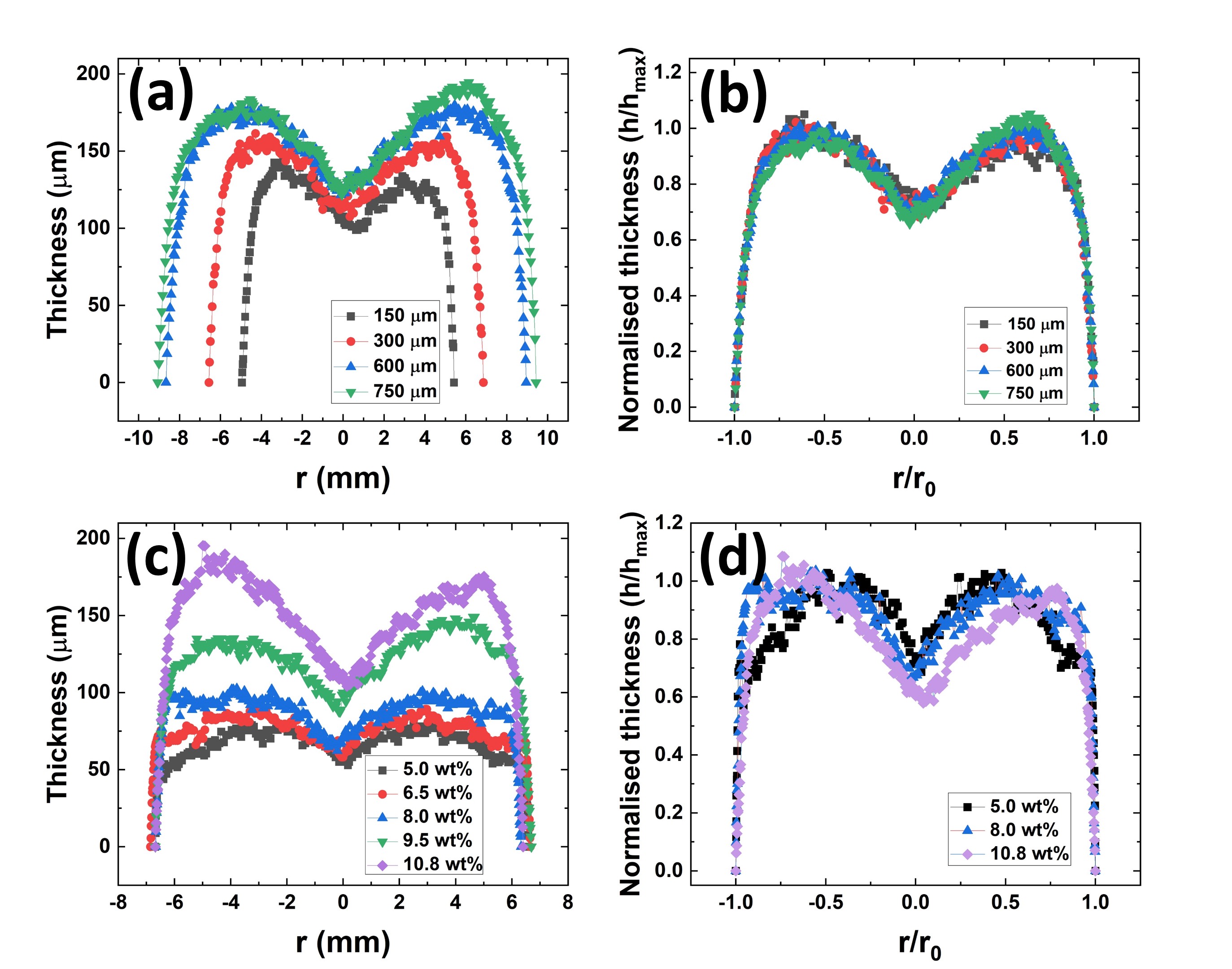}
    \centering
    \caption{(a) The thickness profiles along the diameter of starch films
    formed from different volume of starch solution droplets 
    with 9.5~wt\% initial starch concentration. (b) The corresponding
    normalized thickness profiles as a function of the normalized 
    radial distance. 
    (c) The thickness profiles along the diameter of starch films
    formed from 300~$\mu$l droplets with different starch concentrations.
    (d) The corresponding
    normalized thickness profiles as a function of the normalized 
    radial distance. 
    }
    \label{fig2}
\end{figure}

The starch solution droplets after drying produce circular
films on the substrate. These circular films have azimuthal symmetry 
with a radial variation of their thickness.
Fig.~\ref{fig2}a shows the thickness profiles 
along the diameter of starch films formed 
after drying of different volume of  
droplets with same initial
starch concentration of 9.5 wt\%.  It is found 
that all these profiles show a sharp increase 
near the edge of the films and a dimple at their center 
giving rise to a "M-shaped" thickness profile. 
The films have a maximum thickness
h$_{max}$ and radius r$_0$ depending on 
the initial volume of the droplet. 
The normalized thickness profiles (h/h$_{max}$) of these 
films when plotted with their normalized radial distance (r/r$_0$), 
collapse to a single curve as shown in 
fig.~\ref{fig2}b. Therefore, the 
scaled thickness profile (h/h$_{max}$) 
is a function of only the scaled radial 
distance (r/r$_0$) when other
experimental parameters remain same.

Fig.~\ref{fig2}c depicts the thickness profiles of starch films 
formed from the same volume of droplets (300 $\mu$l) but 
with different initial starch concentrations.
The normalized thickness profiles (h/h$_{max}$) as
a function of normalized radial distance (r/r$_0$) of 
the data shown in fig.~\ref{fig2}c are plotted in 
fig.~\ref{fig2}d. 
The dimple at the center of the film from 
the droplet with 10.8 wt\% initial starch 
concentration is dipper and wider  
compared to that of the film formed 
from the droplet with 5.0 wt\% initial starch concentration.
Therefore, the variation of (h/h$_{max}$) with (r/r$_0$) of the films  
depends on the initial starch concentration
of the droplets.

\begin{figure}[H]    
    \includegraphics[width=8.5cm]{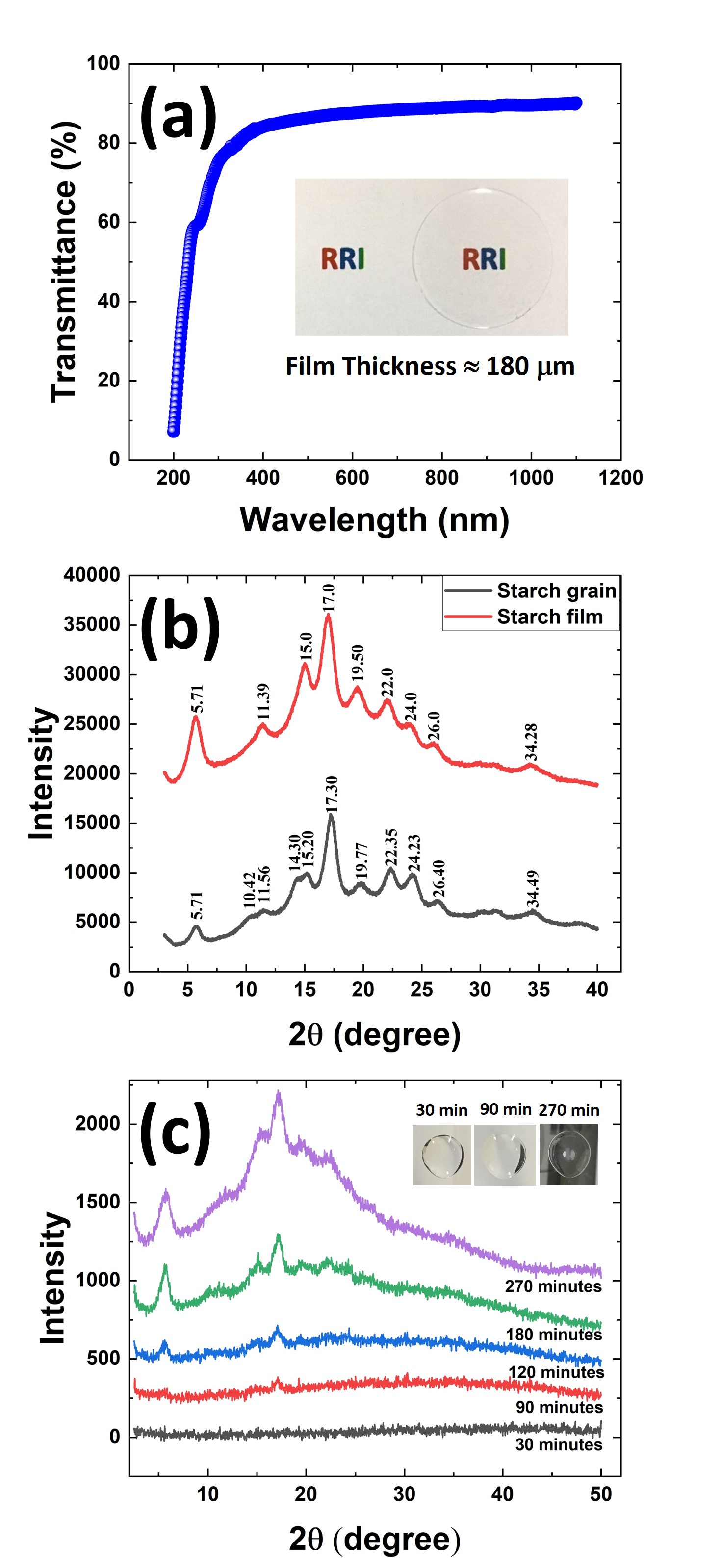}
    \centering
    \caption{(a) The UV-Visible transmission spectrum of a starch film 
    obtained on drying a droplet (600~$\mu$l, 9.5~wt\%). The inset 
    image shows the transparency of this starch film of average 
    thickness about 180~$\mu$m. 
    (b) The XRD intensity profiles of native potato starch grains 
    and the film.
    (c) The XRD profiles at different times after
    dropcasting a 150~$\mu$l droplet with 9.5~wt\% starch
    concentration on a glass plate. The insets show the images of 
    the sample at three different times during drying. The XRD
    intensity profiles are shifted
    vertically for clarity of presentation. 
    }
    \label{fig3}
\end{figure}

The dried starch films are transparent in visible light.
Fig.~\ref{fig3}a shows the UV-Visible transmission spectrum of a 
starch film formed after complete drying of a 600 $\mu$l droplet 
with 9.5 wt\% starch concentration. The inset of fig.~\ref{fig3}a
demonstrates the transparency of this starch film of thickness 
about 180~$\mu$m. The transmittance of this film    
is about 95\% in the wavelength range from 400 nm to 1100 nm and
starts to decrease below 400 nm in the UV range. 

The XRD measurements of starch grains and starch films were performed
to determine their crystalline structure. The XRD profile of 
potato starch grains is shown in fig.~\ref{fig3}b.
The profile exhibits peaks on top of a broad 
amorphous hump. This indicates that the grains are semi-crystalline 
in nature. The peak positions in the XRD profile of 
these grains
are similar to that found for the potato starch grains having B-type of 
crystal polymorph \cite{Jane1997, Cheetham1998, Domene2019}.
This type of crystal polymorph has the hexagonal 
lattice structure with lattice parameter
$a = b = 1.85~nm$, $c = 1.04~nm$ and $\gamma = 120^\circ$ \cite{Imberty1988}. 
The broad hump in this diffraction profile represents
the amorphous regions of the grains coexisting with crystalline regions.
The XRD profile of starch film formed from 600 $\mu$l droplet 
with 9.5 wt\% starch concentration is also shown in fig.~\ref{fig3}b. 
The diffraction peaks on a broad hump in this profile
indicate that the film has semi-crystalline structure 
with coexistence of both crystalline and amorphous regions. 
The similarity of this diffraction profile to that found for
the potato starch grains indicates that the film also has 
B-type of crystal polymorph.

The XRD studies were also carried 
out on a droplet during its drying
on a glass plate to investigate the 
development of the semi-crystalline
structure.
The XRD profiles at different times 
after dropcasting a 150 $\mu$l
droplet with 9.5 wt\% starch concentration are shown 
in fig.~\ref{fig3}c.
It shows that the sample starts to acquire crystalline peaks after 
about 90 minutes when the droplet
begins to recede from its initial edge as shown 
in the inset of fig.~\ref{fig3}c.
The crystalline order perhaps arises from the dried 
starch film starting from the 
peripheral region. 
The SEM studies show that 
the gel structure in droplet of this concentration forms 
almost immediately after its 
preparation. Therefore,  
the gel network formed initially inside the 
droplet does not possess crystalline order.
Further drying increases the 
intensity of both the crystalline peaks 
and the broad hump till complete drying.

\begin{figure}[t!]    
    \includegraphics[width=14cm]{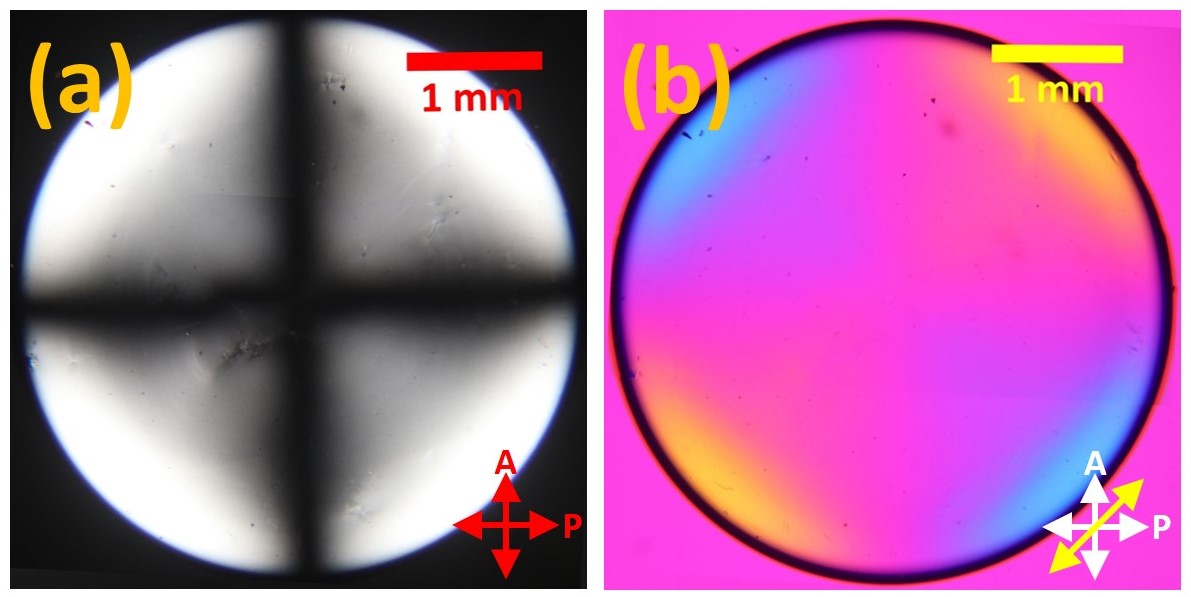}
    \centering
    \caption{The POM images of the starch film (a) between crossed
     polarisers and (b) after inserting of a $\lambda$-plate 
     with the slow axis at an angle of 45$^\circ$ with respect 
     to the polariser.  
    }
    \label{fig4}
\end{figure}

The circular starch films are found to be 
linearly birefringent. Fig.~\ref{fig4}a shows the POM texture of a 
starch film formed after drying a 10 $\mu$l
droplet with 10.8 wt\% of starch concentration.
The film between crossed polarisers appears
gray-white which implies that the retardation introduced by the 
film is in the first order of Levy chart.  
The black brushes of the Maltese cross along 
the pass axis of polarisers 
divide the whole circular 
film into four quadrants and
remain invariant on rotating 
the sample on the microscope stage.
These observations indicate that 
the principal axes of index ellipse on the film are along the 
radial and azimuthal directions 
respectively. The POM 
studies using a $\lambda$-plate 
(530 nm) were performed to determine 
the orientation of the major axis. 
The introduction of the $\lambda$-plate 
with the slow axis at 
an angle of 45$^\circ$ with respect 
to the polariser changes the 
colour of first and third quadrants 
to yellow whereas second and fourth 
quadrants become blue  as shown in 
fig.~\ref{fig4}b. These colours belong 
to the first and second order of Levy 
chart respectively. Therefore the effective 
addition and subtraction of the optical path 
differences occur in the blue and yellow
coloured quadrants respectively. 
These observations imply that the
major and minor axes of the index ellipse on the 
sample are along the azimuthal
and the radial directions of the film respectively.

\begin{figure}[t!]    
    \includegraphics[width=12cm]{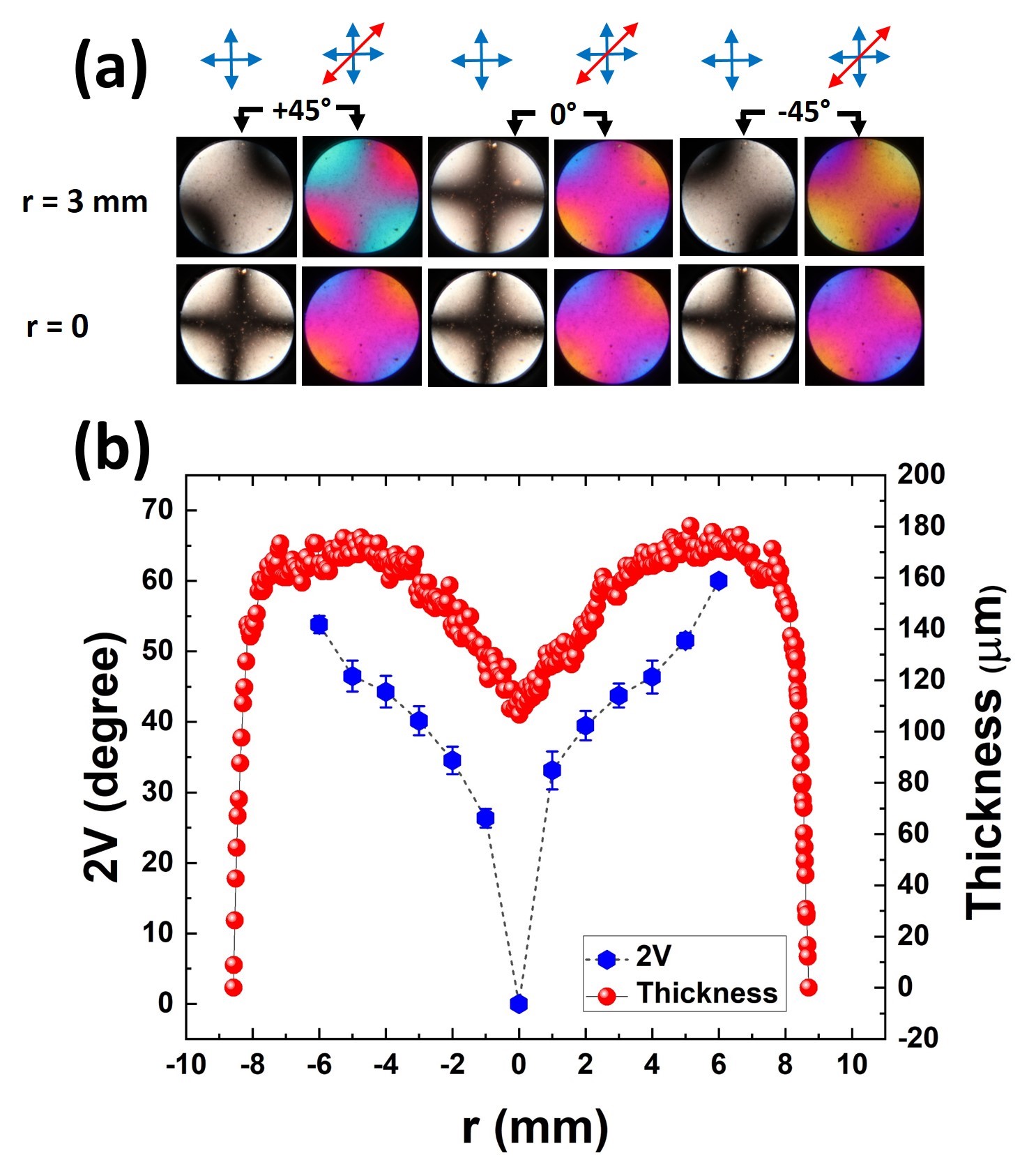}
    \centering
    \caption{(a) The conoscopic figures of starch film at r~$=0$ 
    (bottom row) and at r~$=3$~mm (top row) for three different 
    orientations of the sample on microscope stage with and without 
    the $\lambda$-plate.  
    (b) The variation of the acute axial angle $2V$  
    along the radial direction of a film formed on drying of a
    droplet (600~$\mu$l, 9.5~wt\%). 
    The thickness profile of the film is also shown in this figure.}
    \label{fig5}
\end{figure}

The center of the film between crossed polarisers appears dark 
which implies that this region is either optically
isotropic or uniaxial with an optic axis along the normal to the film.
The conoscopic studies were performed
to determine the nature of the 
optical anisotropy of the dried starch films.
Fig.~\ref{fig5}a shows the conoscopic figures at the center (r $=0$) 
and r $=3$ mm of a starch film 
of diameter about 16~mm formed after drying a 600 $\mu$l droplet
of concentration 9.5 wt\%. 
The bottom row in fig.~\ref{fig5}a shows the conoscopic figures
at the center of the film for three 
different orientations of the sample
on the microscope rotation stage in the absence and presence of a 
$\lambda$-plate in the optical path.    
The intersection point of the crossed isogyres parallel to 
the polarisers in the conoscopic figures without 
$\lambda$-plate is always at the center of the field of 
view. This observation confirms the optical uniaxiality
of the central region with an optic axis normal to the film \cite{pichler1997}.
The insertion of 
the $\lambda$-plate into the optical path changed the colour of the 
first and third quadrants
to yellow whereas the second and fourth quadrants 
became blue (see bottom row of fig.~\ref{fig5}a).
It can be concluded from these observations that the sign of 
the birefringence at the central region of the film is negative.  
Hence, the conoscopic studies confirm that the index ellipsoid
at the center of the film has uniaxial oblate shape.

The top row in fig.~\ref{fig5}a shows 
the conoscopic figures at r $=3$ mm of the film for three different 
orientations of the sample on the microscope rotation stage. 
On rotating the film 
by 360$^\circ$ on the microscope stage, the conoscopic figures 
at this location show that   
the isogyres become crossed at four distinct positions 
with a difference of 90$^\circ$. At $\pm 45^\circ$ with respect to 
a crossed position, the isogyres separate and become 
hyperbolic (see the top row of fig.~\ref{fig5}a) 
which indicates the optical biaxiality of the 
film at this position \cite{pichler1997}.
These hyperbolic isogyres containing the
poles of the two optic axes at their apex are  
centered at the middle of the field of view.
Further, the intersection point of the
isogyres at crossed positions also lies at the center of the field of view.
These observations indicate that one of the 
principal axis of the index ellipsoid with the refractive index 
denoted as $\alpha$
and the optic plane are perpendicular to the film at this biaxial region.  
The orthoscopic studies as discussed above 
confirm that the other principal axes with indices denoted as 
$\gamma$ and $\beta$ on the plane of film are along the azimuthal 
and the radial directions respectively.
The conoscopic studies after 
inserting a $\lambda$-plate (530~nm) 
in the optical path were performed to determine the order
of the principal indices $\alpha$, $\beta$ and 
$\gamma$ at the biaxial region of the film.
The insertion of the $\lambda$-plate for
$+45^\circ$ orientation of the optic plane 
to the polariser changed the colour 
in the region between the uncrossed isogyres 
to blue whereas it became yellow at $-45^\circ$ orientation 
of the optic plane as shown in fig.~\ref{fig5}a (top row). 
Therefore, from these observations it can be concluded that
the film in its biaxial region 
is optically negative and $\alpha$ is the minor principal index.
So the optic plane containing the major principal index $\gamma$ and 
minor principal index $\alpha$ 
lies along the azimuthal direction and perpendicular to the film. 
The intermediate principal axis with the index $\beta$ is along 
the radial direction of the film.

The film is uniaxial at its center and becomes increasingly biaxial
along the radially outward direction. 
The acute axial angle $2V$ between the two optic axes can be 
measured from the conoscopic studies 
(see the supporting information for the 
detail procedure). The variation of
the angle $2V$ along the diameter of
a starch film formed from 600 $\mu$l solution droplet 
with 9.5 wt\% initial starch concentration is shown in 
fig~\ref{fig5}b. 
The angle $2V$ is zero at the uniaxial center 
and it increases monotonically towards the edge of the film.
The hyperbolic isogyres go out 
of the field of view for r $>6$ mm of the film 
(see fig.~S6 in supporting information) and the measurement 
of the axial angle was not possible.

\begin{figure}[t!]    
    \includegraphics[width=8cm]{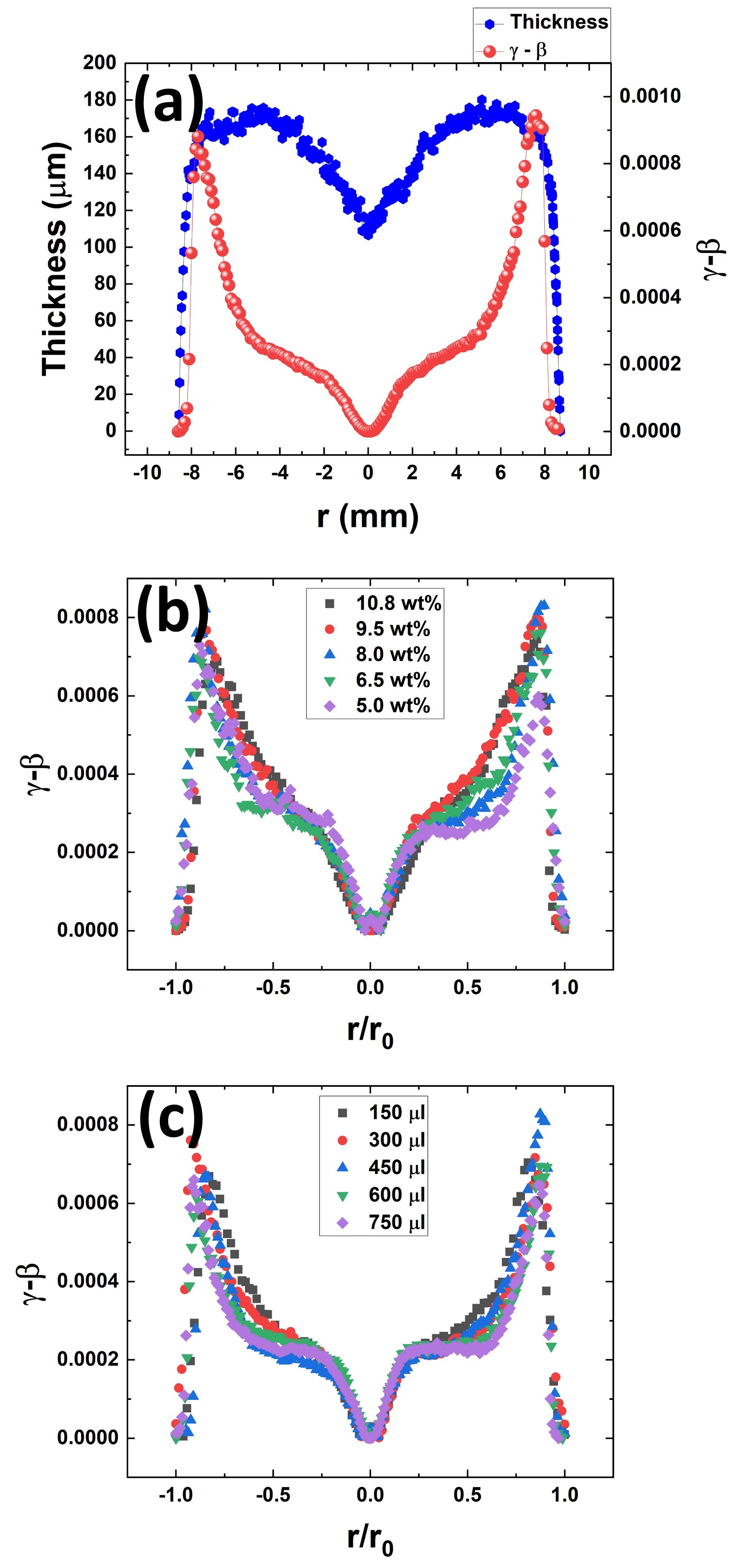}
    \centering
    \caption{(a) The variation of ($\gamma - \beta$) and thickness 
    along the diameter of a starch film obtained 
    from a droplet (600~$\mu$l, 
    9.5~wt\%). The variation of ($\gamma - \beta$) 
    as a function of normalized radial distance on starch films 
    formed from (b) 300~$\mu$l droplets with different starch
    concentrations and (c) different volumes of droplets with 6.5~wt\% 
    starch concentration.}
    \label{fig6}
\end{figure}

The anisotropic optical properties
of these starch films arise due to the 
orientational order of starch bio-polymers 
developed during drying. Therefore,
measuring the linear birefringence along the radial direction of 
starch films provides valuable insight on the structure of 
these films. Fig.~\ref{fig6}a shows the variation of effective 
linear birefringence ($\gamma - \beta$) along
the diameter of a starch film formed after drying
a 600 $\mu$l droplet of concentration 9.5 wt\% 
measured by using the PEM technique. 
This figure also shows the corresponding thickness profile
of the film along its diameter.
The birefringence profile along the radial direction
from the center to the edge 
of the film can be divided into 
three regions with three different slopes.
The measured birefringence at the center of the film is zero
due to its uniaxial nature. 
Initially the birefringence increases rapidly along the radially
outward direction from the center.
Then it increases with a relatively lower slope and again increases
rapidly to the maximum value near the edge of the film.

The birefringence profiles of starch films obtained on drying
the same volume of droplets but with different 
starch concentrations were measured to investigate the effects
of initial starch concentration. 
Fig.~\ref{fig6}b shows the birefringence profiles of films 
obtained on drying 300~$\mu$l droplets with different 
starch concentrations. These profiles are plotted as a function
of the normalized radial distance in the films.
The profiles have similar characteristic
variation  with three different 
slopes along the radial direction.  
But the slope in the middle part 
of the birefringence profile decreases
with decreasing the initial starch concentration.
The maximum values of birefringence near the 
edge of these films are close to each 
other and hence it is independent of the 
initial starch concentration of droplets.

The birefringence profiles of the starch films 
formed after drying different volumes of droplets but with same 
initial starch concentration (6.5 wt\%)
were measured to study the effect of initial droplet volume.
These profiles as a function of the normalized 
radial distance (r/r$_0$) are shown in fig.~\ref{fig6}c.
The superposition of all these data 
indicates that the effective birefringence
($\gamma - \beta$) is a function of (r/r$_0$) only for the other
experimental conditions remaining the same.
The very similar values of maximum 
birefringence near the edge of these films
implies that it does not depend on the 
initial volume of the droplets.

\begin{figure}[H]    
    \includegraphics[width=15cm]{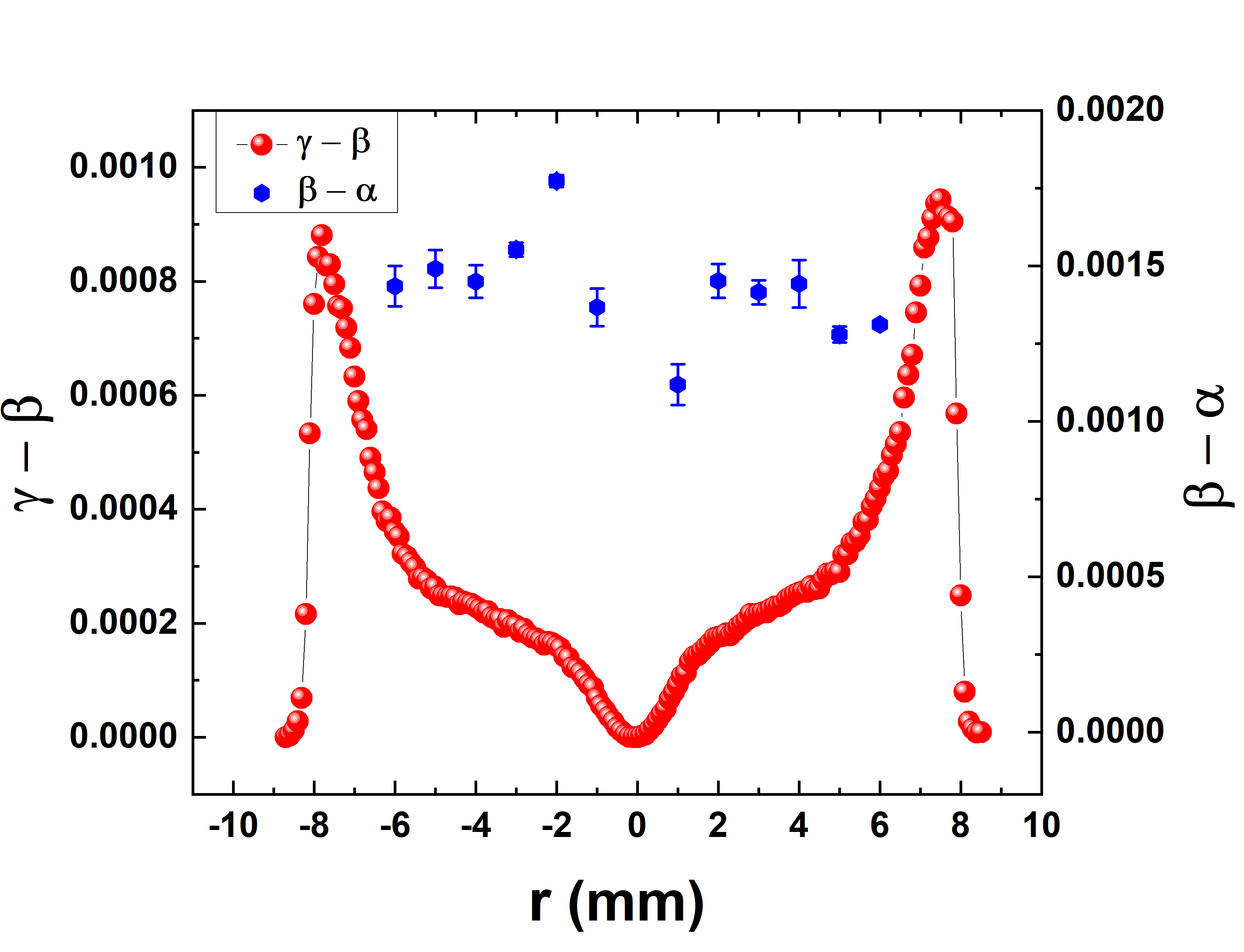}
    \centering
    \caption{ The variation of ($\gamma - \beta$) and 
    ($\beta - \alpha$)
    along the diameter of a starch film obtained on drying of 600~$\mu$l 
    droplet with 9.5~wt\% starch concentration.}
    \label{fig7}
\end{figure}

The variation of ($\beta - \alpha$) along a diameter of a starch 
film can be calculated from the measured values of acute axial 
angle $2V$ and the corresponding values of ($\gamma - \beta$)
(see supporting information for more detail).
Fig.~\ref{fig7} shows the variation of 
($\beta - \alpha$) for a film obtained on drying a
droplet (600 $\mu$l, 9.5 wt\%) which is calculated from the measured 
values of $2V$ (see fig.~\ref{fig5}b) and the values of 
($\gamma - \beta$) also shown in fig.~\ref{fig7}.
It is found that although ($\gamma - \beta$) varies
considerably along the radial direction of the film but 
($\beta - \alpha$) remains almost constant.

\begin{figure}[H]    
    \includegraphics[width=12cm]{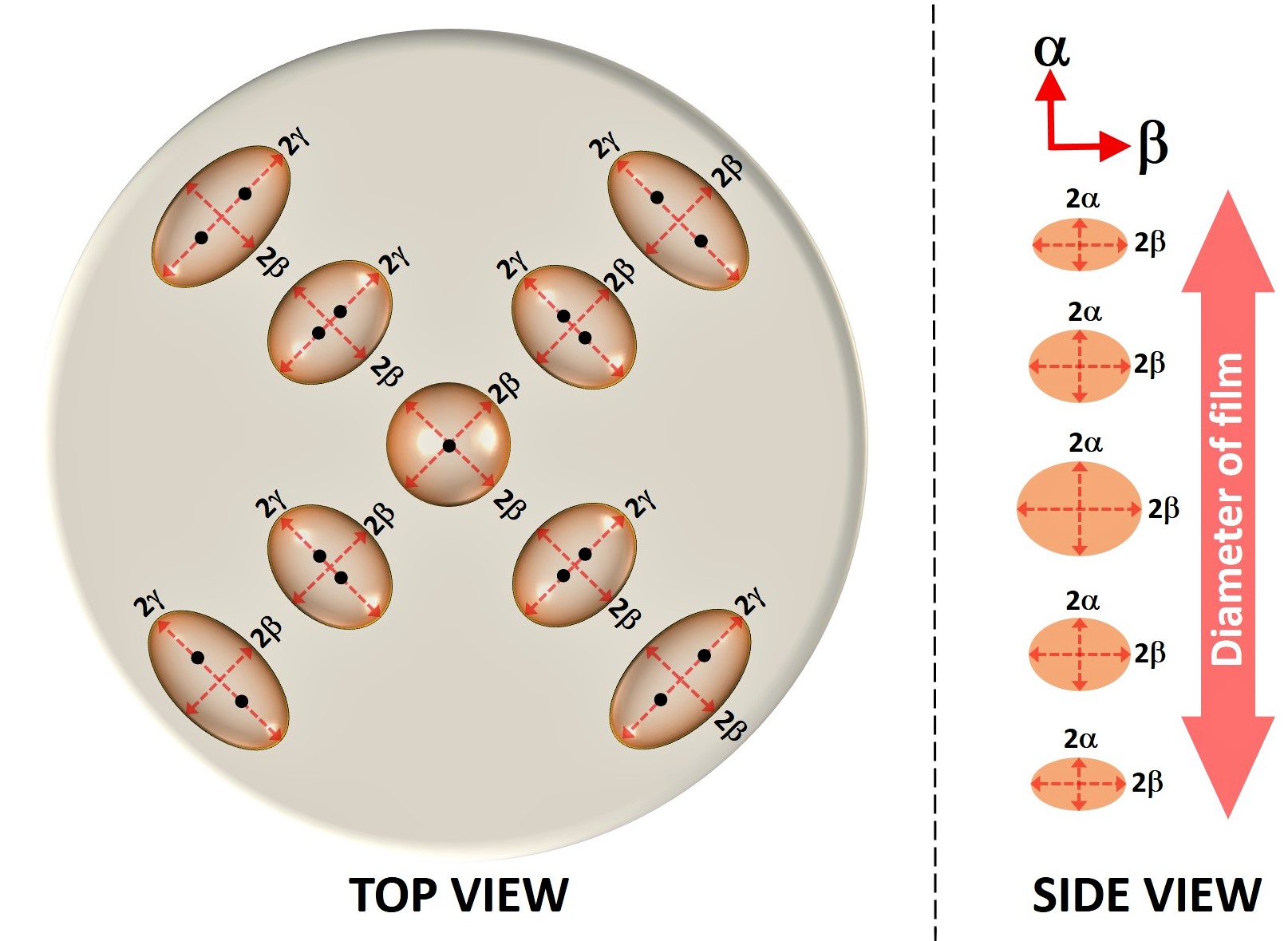}
    \centering
    \caption{[Left] The schematic representation of the index ellipsoids at 
    different positions along the diameter of a starch film. 
    [Right] The variation of index ellipse in $\alpha$-$\beta$ plane along 
    the diameter of the film.}
    \label{fig8}
\end{figure}

The variation of index ellipsoid 
along the diameter of the starch film viewed from the 
top is schematically shown 
in the left side of fig.~\ref{fig8}.
The indicatrix in the biaxial region 
of the film is a general ellipsoid
with three different principal indices $\alpha$, 
$\beta$ and $\gamma$ along the 
normal, radial and azimuthal directions of the film respectively. 
The intersection points between the optic axes 
and the ellipsoid in the biaxial region are 
indicated by a pair of black dots on each 
indicatrix in this diagram.
The increasing angle $2V$ between 
the two optic axes towards the edge 
of the film widens the separation between these
dots along the radially outward direction.
At the center of the film having uniaxial character, the index 
ellipsoid appears circular with the intersection point of 
optic axis at its center. 
The right hand side of fig.~\ref{fig8} shows the variation 
of index ellipse in the $\alpha$-$\beta$ 
plane along the diameter of the film.
At the uniaxial center of the film, the birefringence 
is negative and the index ellipsoid has an oblate shape. 
The indices $\alpha$ and $\beta$ have maximum values
at the center of the film and they
diminish monotonically towards the edge 
keeping their difference almost same.

\begin{figure}[t!]    
    \includegraphics[width=11cm]{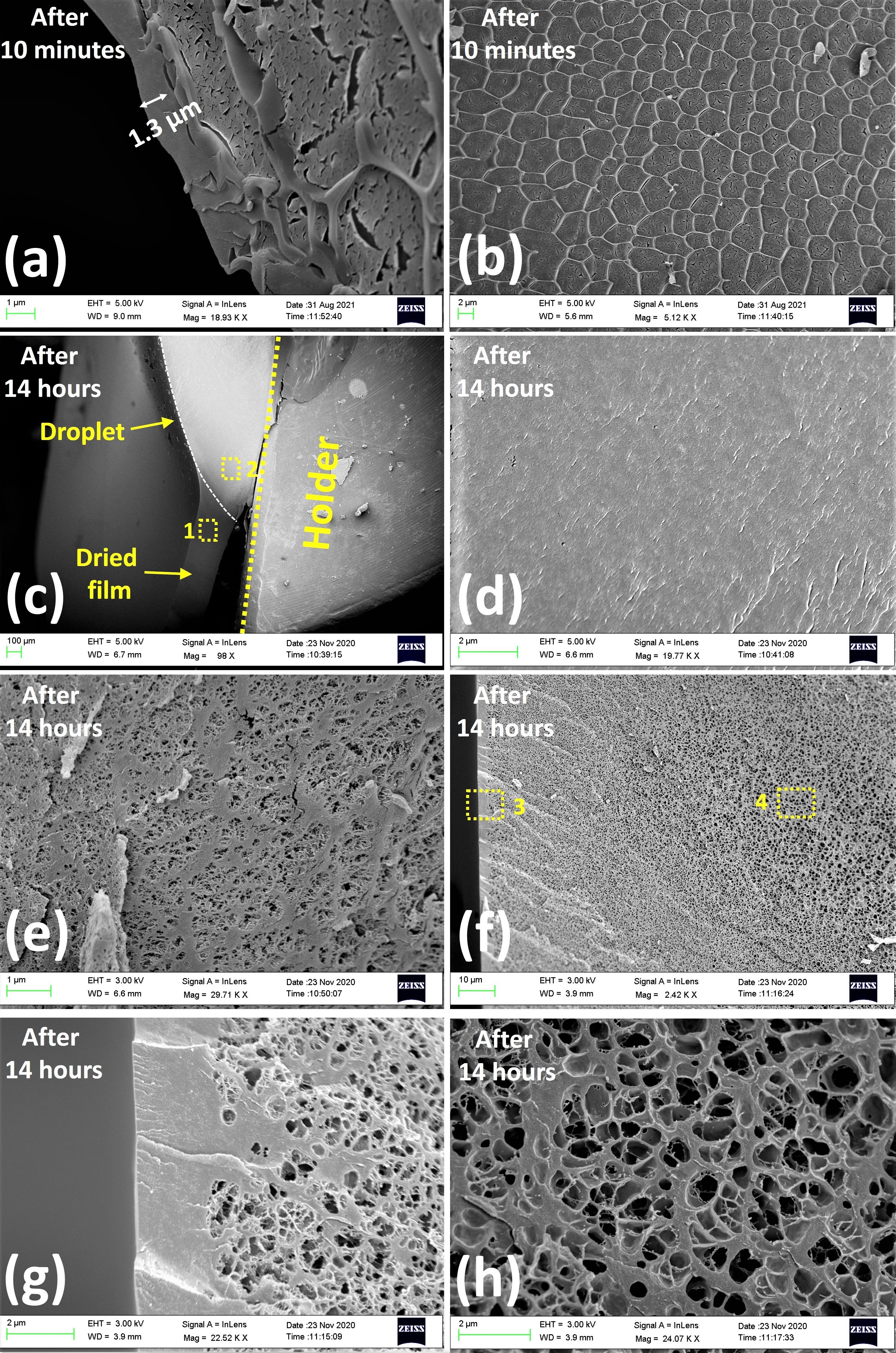}
    \centering
    \caption{The cryo-SEM textures of the vertical cross section of 
    droplets (300~$\mu$l, 9.5~wt\%) at different time 
    after dropcasting. (a) The solid elastic crust on the surface of the 
    droplet after 10~minutes. (b) The cellular 
    network structure inside the droplet after 10~minutes. 
    (c)~The texture near 
    the contact line after 14~hours showing the droplet 
    and the peripheral dried film fixed on the SEM holder. 
    The magnified views of the marked region~1 (d) and  
    region~2 (e) showing the smooth texture of the dried starch film and 
    the mesh network structure inside the droplet
    respectively.  
    (f) The SEM texture near the central interfacial region of the 
    droplet after 14~hours.  The magnified views of the marked 
    region~3 (g) and region~4 (h) showing the crust on top of the droplet 
    and the cellular network structure well inside the droplet
    respectively.}
    \label{fig9}
\end{figure}

The cryo-SEM studies of drying starch droplets were 
carried out to investigate the submicroscopic structures of the sample.
Fig.~\ref{fig9}a,b show the 
cryo-SEM textures of a vertical cross-section of a droplet
(300 $\mu$l, 9.5 wt\%) after 10 minutes of its
dropcasting. The textures clearly show the presence of a solid thin
crust of thickness about 1.3 $\mu$m on the surface 
of the droplet (fig.~\ref{fig9}a) and a cellular 
network inside the droplet (fig.~\ref{fig9}b). 
The solid crust perhaps
forms due to the increase of polymer 
concentration near the free interface of the droplet
during its drying
\cite{doi2006crust, pauchard2003, pauchard20032nd, pauchard20033rd}. 
The cellular network observed inside 
the droplet can account for the gelling
property of the sample. The walls of the cellular 
structures are composed of starch bio-polymers and
starch nano-particles are found to float inside these 
cells (see fig.~S8 in the supporting information). Similar gel network structure
has also been observed during retrogradation of gelatinized starch samples
\cite{wu2012, utrilla2013}.

The circular edge of the droplets (300 $\mu$l, 9.5 wt\%) 
beyond about 7--8 hours starts to recede 
from the initial pinned edge
towards the center with the formation of a peripheral film. 
Fig.~\ref{fig9}c shows the SEM texture of 
vertical cross section of such a droplet
after 14 hours of dropcasting near the receded contact line. 
The regions marked 1 and 2 in this figure correspond to the dried
film outside and the gel just inside the droplet edge respectively.
The magnified view of region 1 as shown in fig.~\ref{fig9}d is 
depicting the smooth texture of the film. 
A mesh network structure made of filaments 
of starch bio-polymers is observed in
the magnified view of region 2 as shown in fig.~\ref{fig9}e. 
This mesh network 
of starch filaments ultimately gives rise to the smooth 
textured film after complete drying.

\begin{figure}[H]    
    \includegraphics[width=14cm]{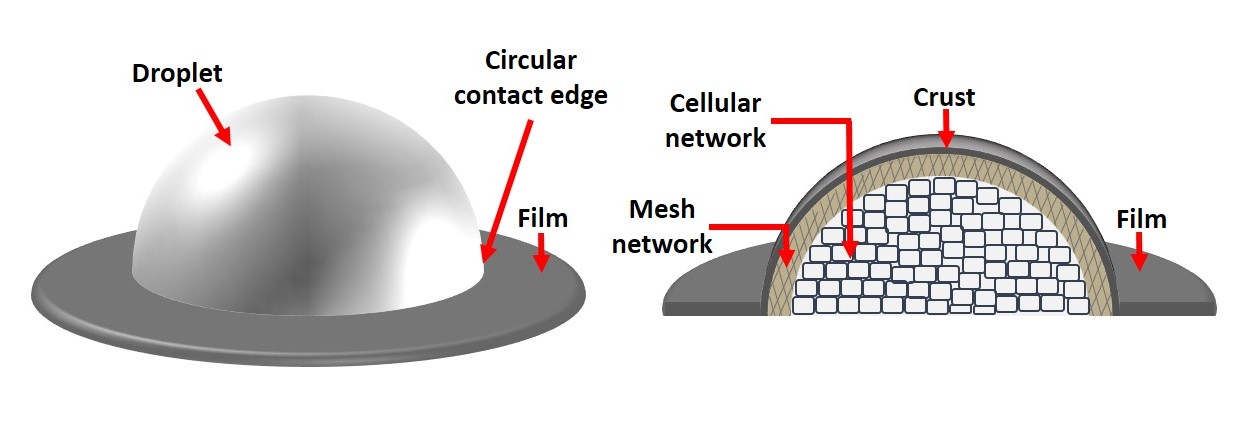}
    \centering
    \caption{ The schematic diagrams of the drying 
    droplet and  its vertical cross section.}
    \label{fig10}
\end{figure}

Fig.~\ref{fig9}f shows the 
cross sectional SEM texture near the free interface of this 
droplet at its central region.
Three different regions with different 
textures can be observed in this figure. 
The magnified view of the marked region 3 near the interface 
is shown in fig.~\ref{fig9}g.
The image shows the smooth texture of a dense crust 
which is same as that found 
for the dried peripheral film shown in fig.~\ref{fig9}d.
Below the crust, 
a mesh network region is developed by the filaments of
starch bio-polymers. The width of this region is 
large compared to that of the crust and has relatively lesser
density than the crust.
A cellular network structure with lowest density is observed below 
the mesh network region. 
The magnified view of the marked 
region 4 of this cellular structure is 
shown in fig.~\ref{fig9}h which clearly
reveals the membranes separating the cells.
Therefore it can be inferred from 
the SEM studies that the drying droplet
develops a relatively thinner solid 
elastic crust at the interface, an intermediate mesh network region 
below the crust and an inner core of 
cellular network structure.
The schematic diagrams of the drying starch droplet and 
its vertical cross section are depicted in fig.~\ref{fig10}.
It is apparent that the cellular network structure gives rise to 
the denser mesh network structure near the droplet interface 
on drying which transforms to the dried 
film at the receding circular
contact edge of the droplet on further drying. 

It has been found that drying a droplet of dilute solution 
of various kind of solutes on a substrate gives rise to a circular 
stain which is known as coffee ring effect\cite{deegan1997, deegan2000}. 
During this drying process, the solute particles 
are driven by the capillary flow towards the edge of the droplet
and deposited there.  
The dried film formed in this way 
shows the highest thickness near its edge. 
Kajiya \textit{et al.} studied this coffee ring effect in the 
deposited film as a function of the concentration 
of a polymeric solution \cite{doi2006}. 
They found that the thickness mismatch of the 
dried film between its rim and the center
reduced with increasing the initial polymer 
concentration and the film became almost flat at a certain 
concentration. Further
increase of the solute concentration 
again started to produce dimple at the middle of the dried film
which became dipper with increasing the solute concentration. 
The mechanism of this dimple formation in 
the films for higher solute concentration 
is not similar to that found in the coffee ring effect.
It is proposed that this surface distortion is  
driven by the crust formation on the droplet 
interface during its drying. 
The similar results of surface distortion in the 
films formed on drying droplets of 
various polymer solutions were also found 
\cite{pauchard2003, pauchard20032nd, pauchard20033rd}.   
In our experiments, a solid crust at the 
free droplet interface and a dense
mesh like network structure below the 
crust are found to form during drying of the droplets. 
The dimple at the center of starch films 
becomes dipper and wider on increasing the 
initial starch concentration as shown in fig.~\ref{fig2}.
Therefore, the "M-shaped" thickness profiles 
of the starch films observed in 
our experiments perhaps can be associated with the 
formation of both the crust and the mesh structure near the droplet interface.
The droplet with higher initial starch concentration perhaps takes 
relatively less time to form the crust and the mesh network 
near the droplet interface leading to the 
formation of a dipper and wider
dimple in the films.

The biaxiality of the dried starch films can be explained
by assuming that the starch filaments forming the mesh network
structure observed experimentally are linearly birefringent. 
These filaments near the 
circular contact line of the drying droplet tend to align 
themselves parallel to this line.
This alignment of the starch filaments perhaps gives rise to the 
biaxial starch film with three principal indices 
$\alpha$, $\beta$ and $\gamma$ along 
normal, radial and azimuthal directions respectively. 
The alignment tendency of these filaments along the azimuthal direction
gives rise to the highest index $\gamma$ along this direction
near the edge of the film. The other two principal directions being 
not equivalent during drying gives rise to the biaxility of 
the film away from the center.  
The gradually diminishing orientational order of 
the filaments decreases the value of the 
principal index $\gamma$ monotonically 
along the radially inward direction of the film.
As a consequence of this, the value 
of other two principal indices 
$\alpha$ and $\beta$ increases 
simultaneously along the same direction 
keeping their difference 
constant as found experimentally.

\section{Conclusion}
We study the drying dynamics of droplets of 
gelatinized potato starch solutions on a flat substrate
at room temperature and humidity. The droplets 
are initially pinned at their circular contact line to the 
substrate. The droplet edge subsequently starts 
to recede towards the center,
leaving a transparent and linearly birefringent 
film from its initial pinned edge. The circular 
starch films formed after complete drying of droplets   
are azimuthally symmetric and show a 
"M-shaped" thickness profile along their diameter. These 
films are partially crystalline and appear 
transparent in visible light. The POM studies 
show that the films at their center are optically uniaxial with the optic 
axis perpendicular to the films but biaxial 
away from the center. In the biaxial region, the 
three principal indices $\alpha$, $\beta$ and $\gamma$ 
are along the normal, radial and azimuthal directions 
of the film respectively with $\alpha < \beta < \gamma$. Both the 
uniaxial and biaxial regions of the films are found to be
optically negative. The effective 
birefrigence ($\gamma-\beta$) 
increases along the radially outward direction of the films 
whereas $(\beta-\alpha)$ remains almost constant.
The cryo-SEM studies show that the starch solutions 
with time develop a cellular network structure 
giving rise to their gelling property. 
The SEM texture of the vertical cross-section of a drying droplet
shows a cellular network structure at the core, 
an intermediate mesh region and a solid crust as the outer most layer.
The intermediate mesh region consists of filaments of starch bio-polymers.
The alignment of these filaments
near the receding circular edge of the drying droplet perhaps 
gives rise to the linear birefringence of the deposited solid film.

\begin{acknowledgement}

We would like to thank Ms. Vasudha K. N. for her help in acquiring
XRD data and UV-Vis transmission spectra. We also would like to 
thank Mr. K. M. Yatheendran for his help in cryo-SEM imaging.

\end{acknowledgement}

\begin{suppinfo}

The supporting information contains (a) the 
schematic figure of the optical retardation 
measurement setup by using PEM, (b) schematic figure of a cryo-SEM holder,
(c) the weight loss curve of a starch 
solution droplet (150 $\mu$l, 9.5 wt\%) during its drying 
on a plastic substrate,  
(d) the conoscopic figures along 
the radial direction of a starch film 
formed after drying a droplet (600 $\mu$l, 9.5 wt\%), 
(e) the cryo-SEM texture of a starch 
solution droplet (300 $\mu$l, 9.5 wt\%) 
after about 10 minutes of its dropcasting and (f) the 
derivation of the formulas for measuring the acute axial 
angle $2V$ and the effective birefringence ($\beta-\alpha$). 

The following files are available free of charge.
\begin{itemize}
  \item Filename: Supporting information
\end{itemize}

\end{suppinfo}

\bibliography{bibfile}

\end{document}


\begin{center}
\textbf{\underline{\large{Supporting information}}}

\vspace{10mm}

\textbf{\Large{Optical and structural properties of starch
films formed on drying droplets of gelatinized
starch solutions}}

\vspace{8mm}

\textbf{Subhadip Ghosh$^a$, Arun Roy$^a$}

\vspace{5mm}

$^a$Raman Research Institute, C.V. Raman Avenue, Sadashivanagar, Bangalore 560080, India.

\end{center}

\begin{figure}[h!]    
    \includegraphics[width=17cm]{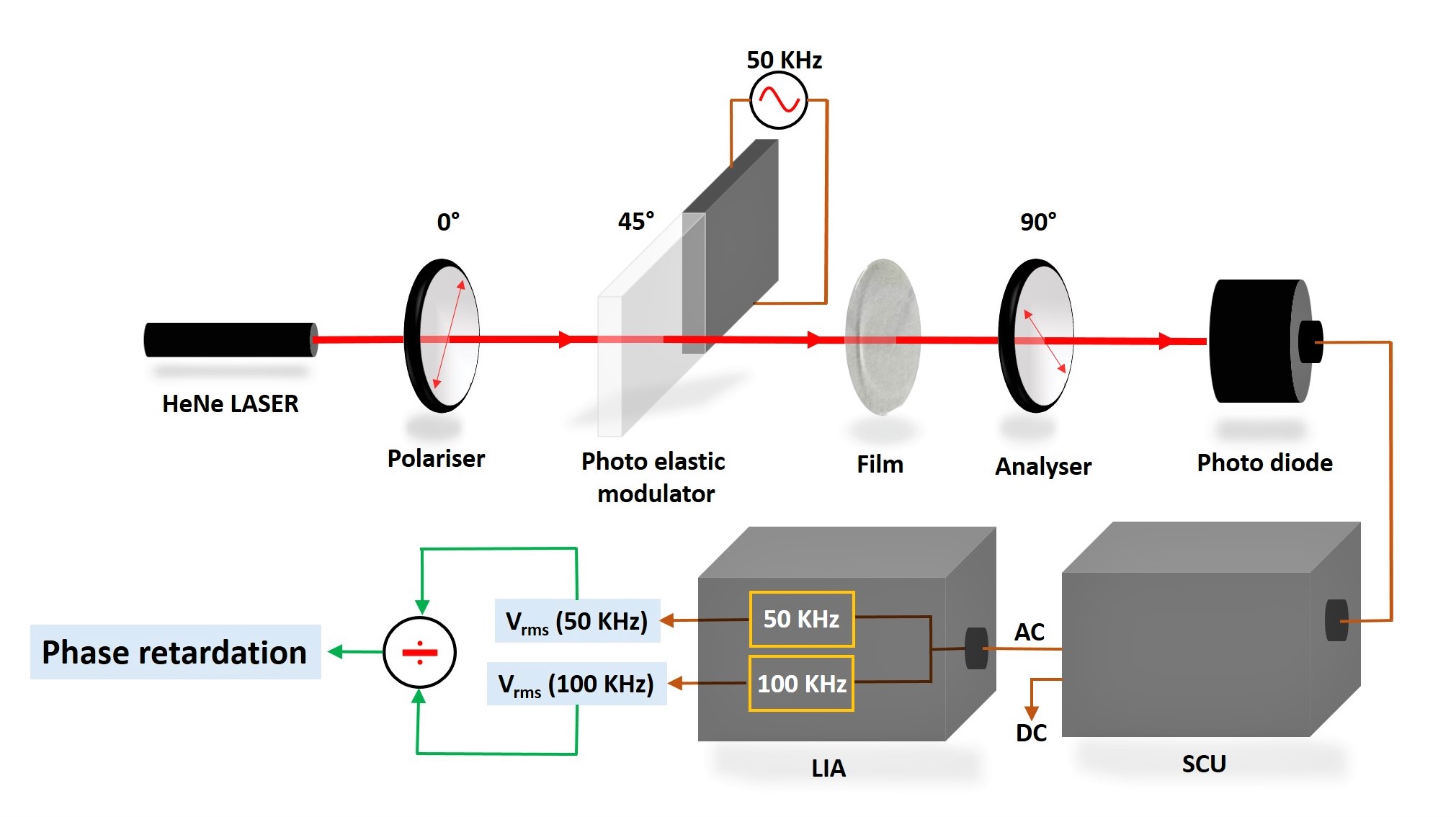}
    \centering
    \caption{Schematic figure of the optical retardation measurement setup.}
    \label{LBS}
\end{figure}

\begin{figure}[h!]    
    \includegraphics[width=11cm]{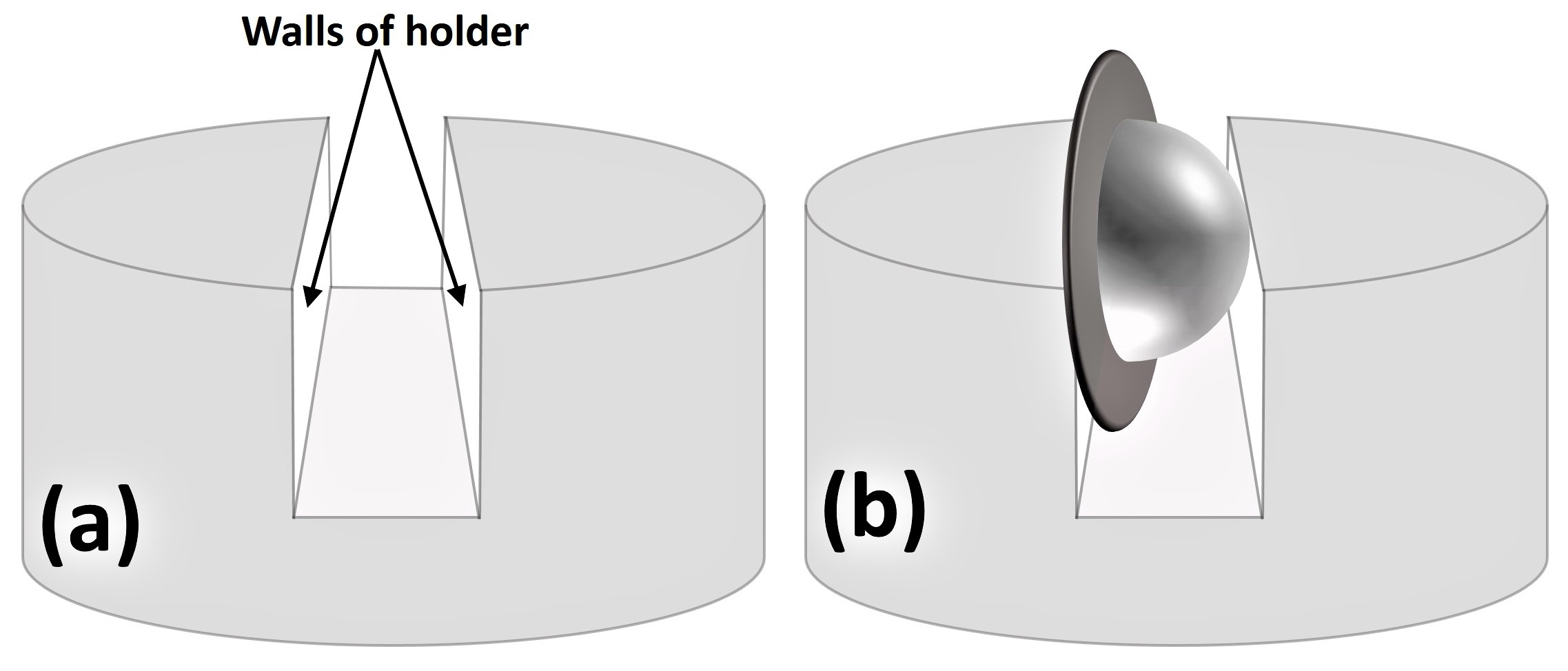}
    \centering
    \caption{Schematic figures of (a) the sample holder for cryo-SEM
     studies  and (b) the sample fixed vertically on it.}
    \label{HOLDER}
\end{figure}

\begin{figure}[H]    
    \includegraphics[width=14cm]{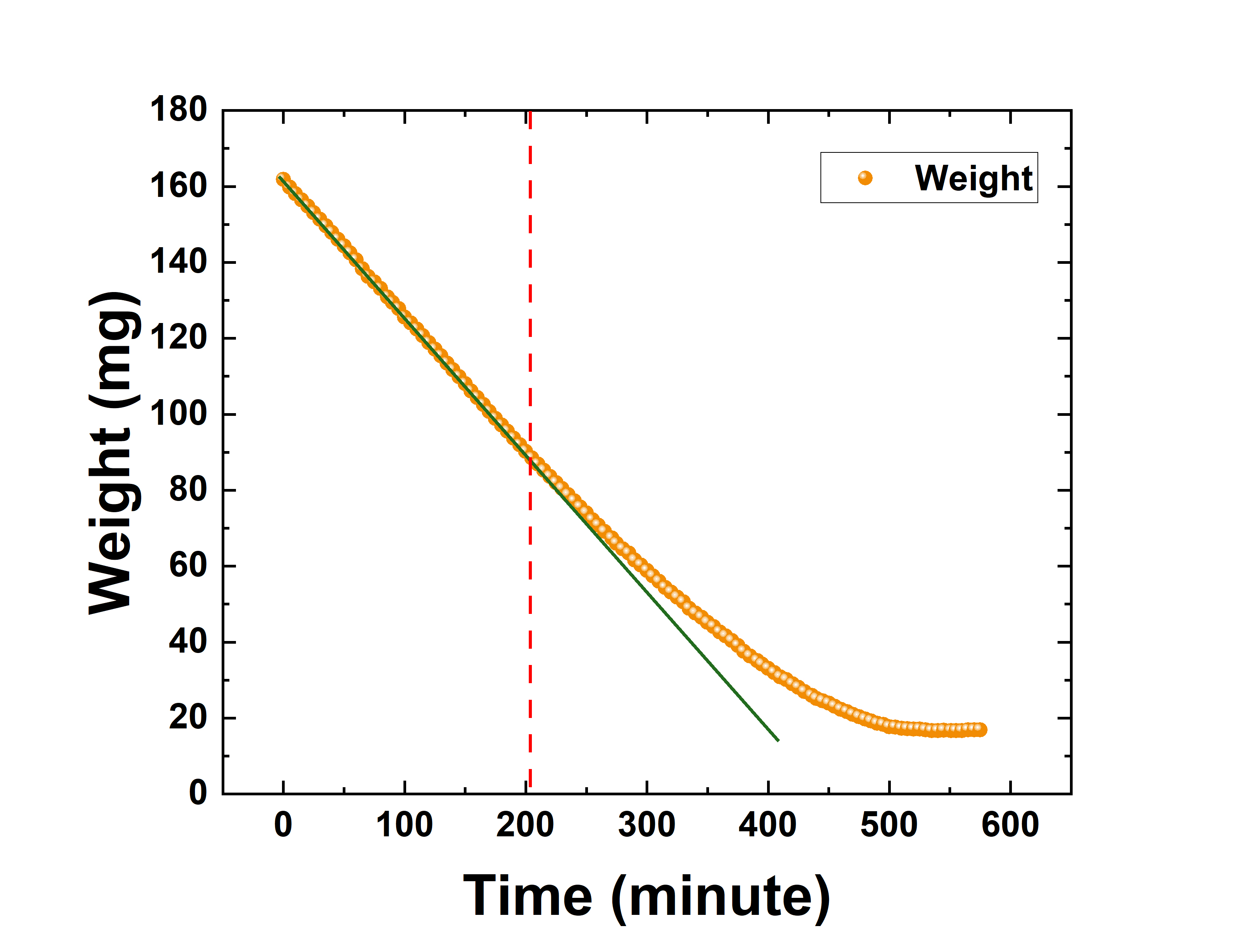}
    \centering
    \caption{The variation of weight during drying 
    of a 150~$\mu$l starch solution droplet with 9.5~wt\% of initial 
    starch concentration. The green straight line shows the initial 
    linear variation of weight. }
    \label{Drying_drop}
\end{figure}

\section{\textbf{The derivation of the formula for 
measuring the acute axial angle}}
\begin{figure}[H]    
    \includegraphics[width=14cm]{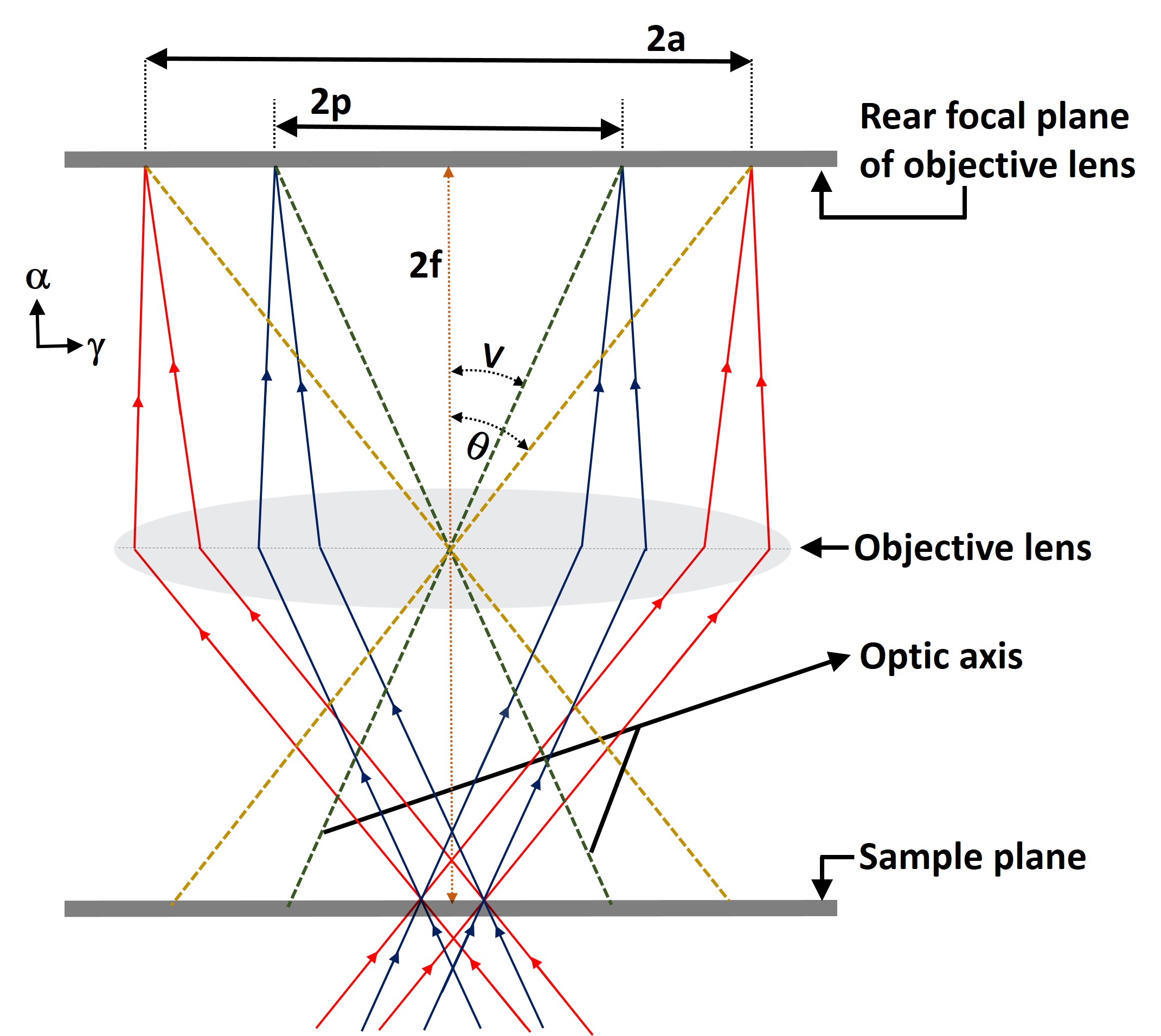}
    \centering
    \caption{Schematic ray diagram of the conoscopic measurement setup.}
    \label{CON_ANG_2}
\end{figure}

The two optic axes in the biaxial region of starch film 
lie on the $\alpha$-$\gamma$ plane where $\alpha$ 
and $\gamma$ are the minor and major principal indices which 
are along the normal and azimuthal directions of the film respectively. 
The acute axial angle $2V$ between the two optic axes 
was calculated from the conoscopic figures. Fig.~\ref{CON_ANG_2}
shows the schematic ray diagram of the conoscopic measurement setup.
The diagram depicts the light rays in the
$\alpha$-$\gamma$ plane of the sample containing both the 
optic axes. In this setting, the light rays from 
the illuminated part of the sample 
kept in the front focal plane of the objective lens are focused 
on the rear focal plane. Hence from fig.~\ref{CON_ANG_2},
it can be written as
\begin{equation}
\frac{a}{f} = \tan \theta
\label{equation:tantheta}
\end{equation}
and,
\begin{equation}
\frac{p}{f} = \tan V
\label{equation:tanV}
\end{equation}
From eqn.~\ref{equation:tantheta} and eqn.~\ref{equation:tanV}, one can 
write,
\begin{equation}
\frac{p}{a} = \frac{\tan V}{\tan \theta} = 
\frac{\sin V}{\cos V} \times \frac{\cos \theta}{\sin \theta}
\end{equation} 
Since the numerical aperture ($NA$ = $n \times \sin \theta$) of
the 50X objective lens used in our experiments is 0.5 
and the refractive index for the intervening
air medium $n =1.0$, the value of $\theta=30^\circ$ in our setup. 
Then for $V$ between 0$^\circ$ and 30$^\circ$,
$(\cos \theta/\cos V) \approx 1$ and we can write 
\begin{equation}
\label{eqn:conoscopy}
\boxed{\sin V = \frac{NA \times p}{a} = 
\frac{NA \times 2p}{2a} }
\end{equation}

\begin{figure}[H]    
    \includegraphics[width=8cm]{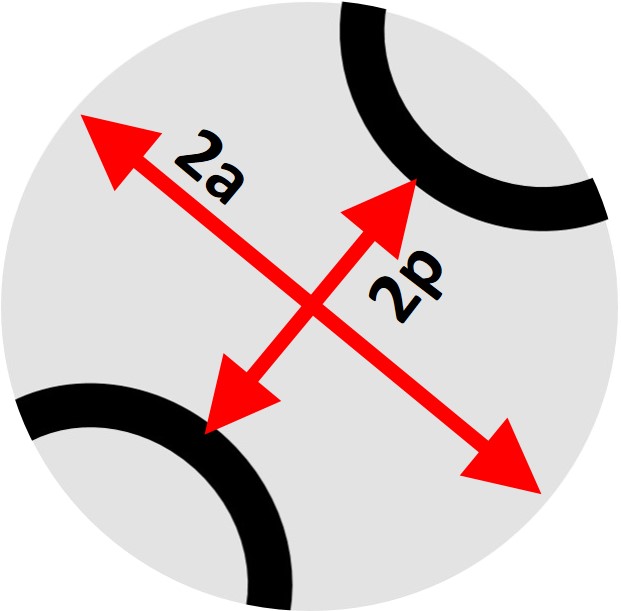}
    \centering
    \caption{A schematic conoscopic figure 
    having two uncrossed hyperbolic isogyres and also depicting
    the distances $2p$ and $2a$. }
    \label{CON_ANG}
\end{figure}

Fig.~\ref{CON_ANG} shows the schematic 
conoscopic figure of a biaxial region of the 
film when the optic plane makes an angle of 45$^\circ$ to the polariser. 
The $2p$ and $2a$ are the distance between the 
poles of the optic axes and the
diameter of the circular field of view respectively. The acute axial angle $2V$
between the optic axes can be determined from eqn.~\ref{eqn:conoscopy} 
by measuring the ratio of $2p$ and $2a$.

\begin{figure}[H]    
    \includegraphics[width=14cm]{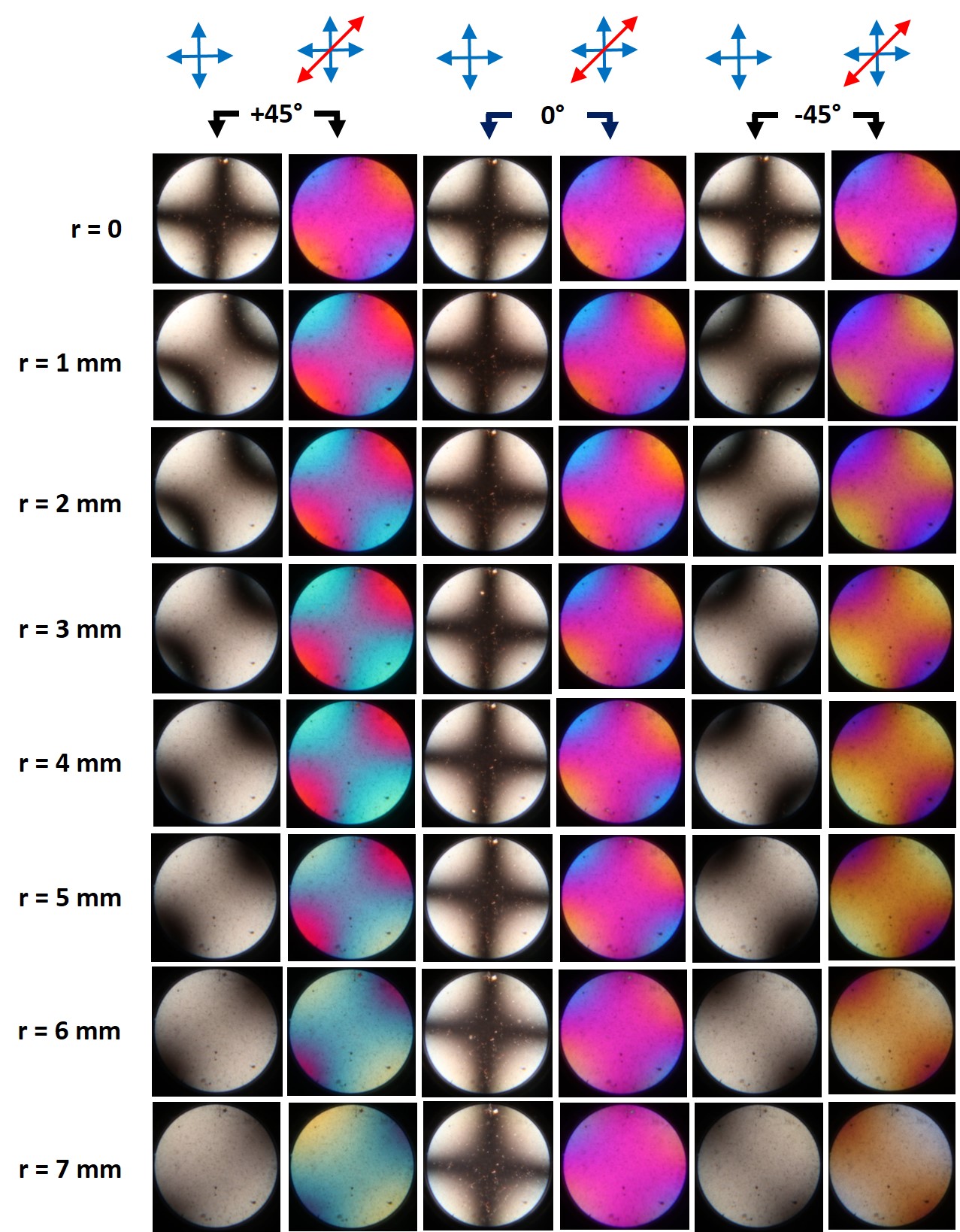}
    \centering
    \caption{Conoscopic figures at different positions along 
    the radial direction of a starch film of
    radius about 8 mm formed from a droplet (600 $\mu$l, 9.5 wt\%). The
    distance between two consecutive positions of measurement is 1 mm. }
    \label{Conoscopy}
\end{figure}

The conoscopic figures at different positions along the radial direction
of a starch film formed from 600 $\mu$l 
droplet with 9.5 wt\% of initial starch concentration 
are shown in fig.~\ref{Conoscopy}. The measurements were performed
at an interval of 1 mm  along the radial direction.
The angle $2V$ between the two optic 
axes along the diameter of the film was 
calculated from these conoscopic figures 
by using eqn.~\ref{eqn:conoscopy}.

\section{\textbf{Formula for measurement of ($\beta-\alpha$)}}

\begin{figure}[H]    
    \includegraphics[width=10cm]{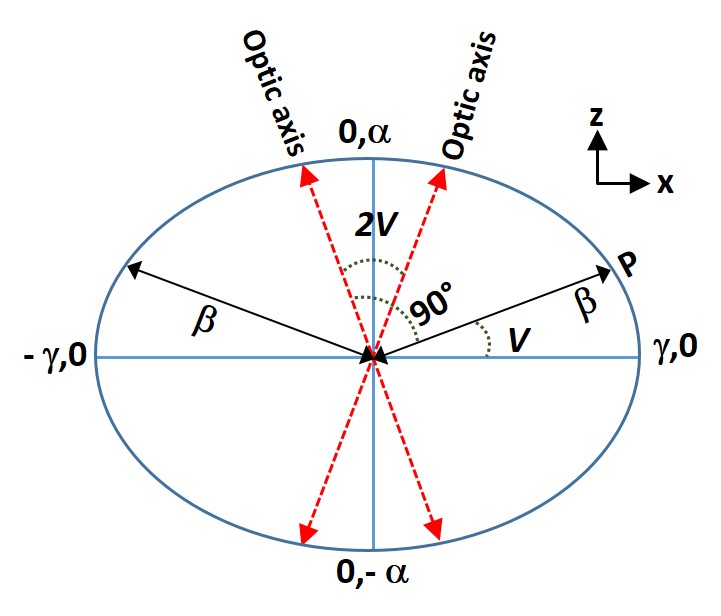}
    \centering
    \caption{The index ellipse in the 
    optic plane of starch film.}
    \label{SCHEM_2}
\end{figure}

In the biaxial region of the circular starch 
film, the three principal indices 
$\alpha$, $\beta$ and $\gamma$ are 
along the normal, radial and 
azimuthal directions respectively where $\alpha<\beta<\gamma$. 
The index ellipse in the optic plane (i.e, $\alpha$-$\gamma$ plane) 
is shown in fig.~\ref{SCHEM_2}. The equation of this
index ellipse can be written as
\begin{equation}
\label{equation:conoscopy1}
\frac{x^2}{\gamma ^2} + \frac{z^2}{\alpha ^2} = 1
\end{equation}
where $\gamma$ and $\alpha$ are the major and minor indices respectively.
The sections of the optical indicatrix 
perpendicular to optic axes are circular with radius $\beta$.
Therefore, the length of the radius vector to the point P on the 
index ellipse in fig.~\ref{SCHEM_2} is the intermediate index $\beta$.
Then using eqn.~\ref{equation:conoscopy1}, 
it can be shown that
\begin{equation}
\label{equation:m2}
\tan^2 V = \frac{1-\frac{\beta ^2}{\gamma ^2}}{\frac{\beta ^2}{\alpha ^2} - 1}
\end{equation}

The principal refractive indices 
$\alpha$, $\beta$ and $\gamma$ of starch 
film are expected to have similar order of magnitude to 
that of water or glass.
The measured value of effective birefringence $(\gamma - \beta)$ of the starch
films is found to be of order $10^{-4}$ which is very small compared to 
the principal values of the indices.  As $60^\circ > 2V > 30^\circ$ 
in most of the biaxial region of the starch films (see fig.~5b in main text), 
it implies that $(\beta - \alpha)$ 
also has similar order of magnitude as that of $(\gamma - \beta)$. 
Therefore, using $(\gamma - \beta)\ll\beta$ and 
$(\beta - \alpha)\ll\beta$, eqn.~\ref{equation:m2}
can be approximated as

\begin{equation}
\boxed{ \tan^2 V = \left[\frac{\gamma - \beta}{\beta - \alpha}\right]}
\end{equation}
or
\begin{equation}
\label{eqn:ba}
\boxed{ (\beta - \alpha) = \left[\frac{\gamma - \beta}{\tan^2 V}\right]}
\end{equation}

\begin{figure}[H]    
    \includegraphics[width=14cm]{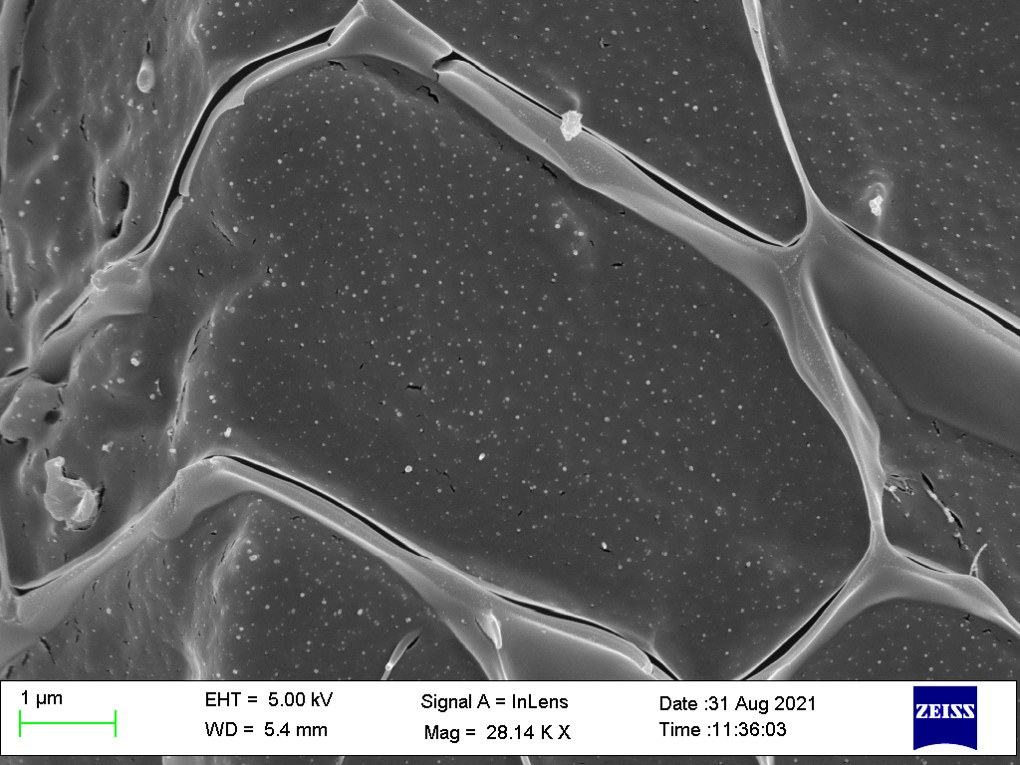}
    \centering
    \caption{The cryo-SEM texture of the cross section 
    of the starch solution droplet with 300 $\mu$l 
    initial volume and 9.5 wt\% initial starch concentration 
    after 10 minutes of its dropcasting on a plastic substrate. 
    The texture shows the starch nano-particles are floating in 
    the solution bounded by the starch membranes.}
\end{figure}